\documentclass[12pt,A4]{article}

\usepackage[utf8]{inputenc}
\usepackage[T1]{fontenc}
\usepackage{times}
\usepackage{mathptmx}
\usepackage[english]{babel}
\usepackage{amsmath}
\usepackage{amsfonts}
\usepackage{amssymb}
\usepackage{makeidx}
\usepackage{graphicx}
\usepackage{hyperref}							
\usepackage{geometry}							
\usepackage{fancyhdr}							
\usepackage[acronym]{glossaries}				
\usepackage{lineno}								
\usepackage{authblk}
\usepackage{sectsty}
\usepackage{setspace}
\usepackage{titlesec}
\usepackage{verbatim}							
\usepackage{floatrow}							
\usepackage{subcaption} 						
\usepackage{cleveref}							
\usepackage{etoolbox}							

\fancypagestyle{plain}{
	\fancyhf{}
	\lhead{\fontsize{8}{10}\selectfont \textit{Tirico et al. Morphogenesis of street networks}}
	\fancyfoot[LO]{\fontsize{8}{10} \selectfont \textit{Book of Abstracts of ECTQG2021, Manchester, United Kingdom, 3 to 5 November 2021}  }
	\fancyfoot[RO]{\fontsize{8}{10} \selectfont \textit{\thepage}\qquad}
}

\makeatletter
\patchcmd{\@maketitle}{\LARGE}{\fontsize{20}{15}\selectfont}{}{}
\makeatother

\title{\textbf{Morphogenesis of street networks. A reaction-diffusion system for self-organized cities}}
\author[1]{\textbf{Michele TIRICO}}
\author[1]{\textbf{Stefan BALEV}}
\author[1]{\textbf{Antoine DUTOT}}
\author[1]{\textbf{Damien OLIVIER}}
\affil[1]{Normandy Univ, UNIHAVRE, LITIS, 76600 Le Havre, France,[name].[surname]@univ-lehavre.fr}

\date{}

\newcommand\myciteauthor[1]{\citeauthor{#1} (\citeyear{#1})}

\newacronym{rd}{RD}{reaction-diffusion}
\newacronym{bc}{BC}{betweenness centrality}
\makeglossaries

\sectionfont{\fontsize{12}{15}\selectfont}
\titlespacing{\section}{0em}{1em}{0em}
\usepackage[style=authoryear-icomp,backend=biber,maxcitenames=1,giveninits=true, date=year, doi=false, isbn=false, url=false,maxbibnames=9,maxcitenames=1]{biblatex}
\AtEveryBibitem{\clearfield{editor}}
\begin{document}
	\maketitle
	
	
	\section{Introduction} 	\label{sec:introduction}
	Morphogenesis is the natural or artificial process that determine physical forms, shape and patterns of a self-organized system and specialize its subpart \autocite{bourgine_morphogenesis_2011,darcy_growth_1917}. 
	Within urban systems, morphogenesis concern the formation of its physical elements (built-up areas, street networks, public spaces) and the specialization of its suburbs (residential, productive, leisure).
	Street networks represent a major organizational component of the urban systems \autocite{marshall_streets_2005}. 
	Streets are the backbone of the city, the structural support of human activities, the physical witness of the evolution of the urban area. 
	Understanding street network evolution is revealing important information about the growth of cities and their functioning.
	Modelling the urban system, simulate realistic scenarios and quantitatively measure properties of results is one of well adapted approaches to study the street network morphogenesis \autocite{pumain_urban_2017}.  
		
	Complexity theory \parencite{morin_introduction_2005} provides a comprehensive framework to study urban morphogenesis and more generally urban dynamics \parencite{portugali_complexity_2006}. 
	A general agreement can be find into the principle that urban systems are characterized by decentralized interactions between their constitutive elements \parencite{batty_new_2013, pumain_villes_1989}. 
	This context is a source of self-organization and it surrounds the emergence of unanticipated and new properties. 
	Spatial patterns can be observed, as like an aggregation of elements, a regularity of properties or a specialization of a part.    
	The system is surrounded by an environment, which exchange with it and play the role of a morphogenetic actor. 
	Forms appear as an overlapping of different processes, becoming the result of opposite actions which favour or inhibit the growth. 
	To understand those phenomena we cannot sum it up with an overall design intent and we must consider them in their wholeness. 
	
	In the following, we present a street network model generator, we proposed an application in an real case (Fécamp, Normandy, France), we measured different scenarios and and we discuss them.
	
	\section{Modelling spatial network morphogenesis}
 	\begin{figure} [t!] 	\def\scaleFig{.8} 	 
		\includegraphics [ width=\scaleFig \linewidth ]{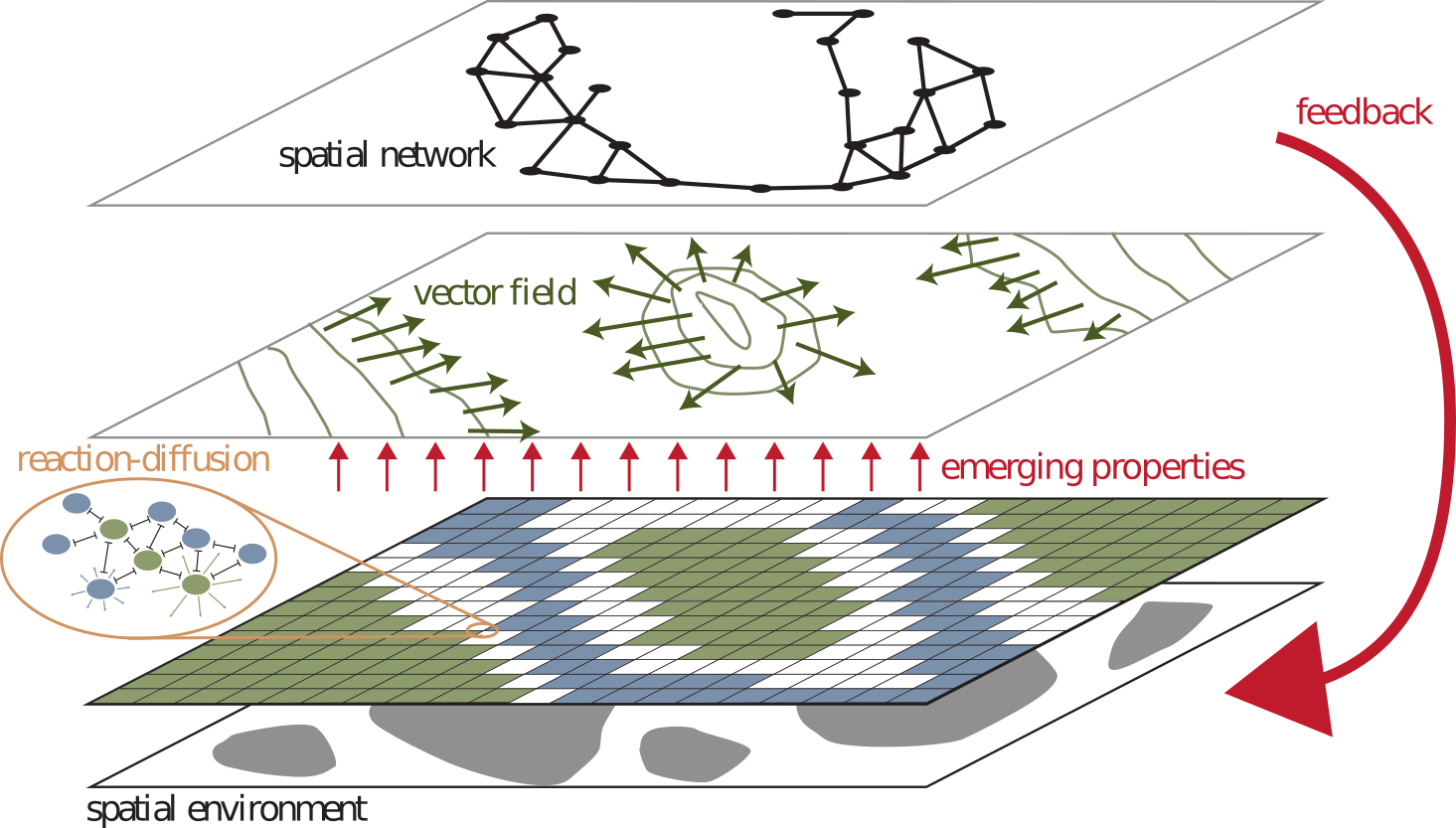}	
		\caption{The concept. Morphogens govern the formation of a network which in turn retro-acts to their behaviour.}
		\label{fig:concept}
	\end{figure}	
	
	\begin{figure} [t] \def\scaleFig{.20} \def\trimleft{4.9} \def\trimbottom{1} \def\trimright{5}  \def\trimtop{1}  \def\hDist{.4cm}
		\centering
		\begin{subfigure}{\scaleFig \textwidth}	\includegraphics[trim= 6cm 1cm 6cm 1cm,clip,width=\textwidth]{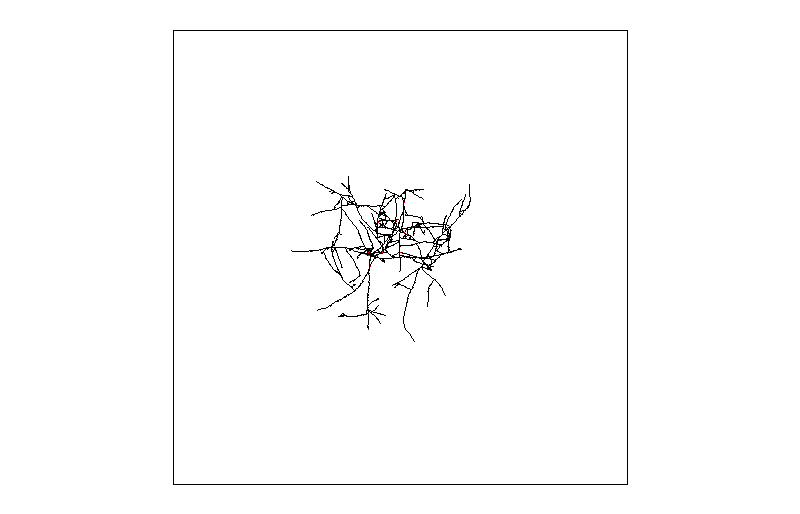}	\end{subfigure}	\hspace{\hDist}	
		\begin{subfigure}{\scaleFig \textwidth}	\includegraphics[trim= \trimleft cm \trimbottom cm \trimright cm \trimtop cm,clip,width=\textwidth]{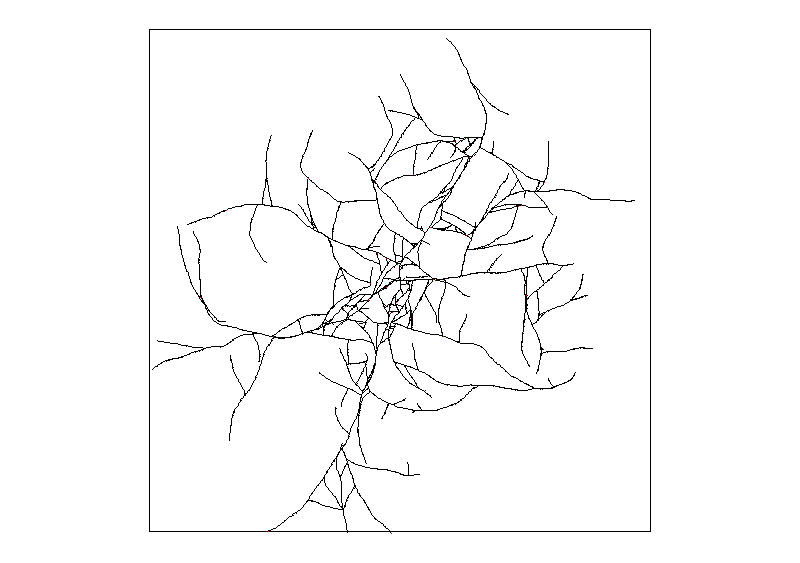}	\end{subfigure}	\hspace{\hDist}	
		\begin{subfigure}{\scaleFig \textwidth}	\includegraphics[trim= \trimleft cm \trimbottom cm \trimright cm \trimtop cm,clip,width= \textwidth]{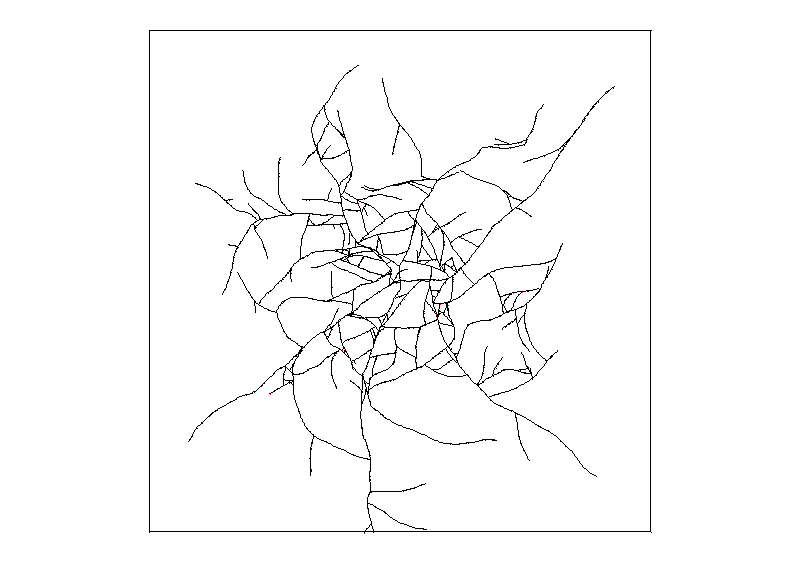}	\end{subfigure} \hspace{\hDist}	
		\begin{subfigure}{\scaleFig \textwidth}	\includegraphics[trim= \trimleft cm \trimbottom cm \trimright cm \trimtop cm,clip,width= \textwidth]{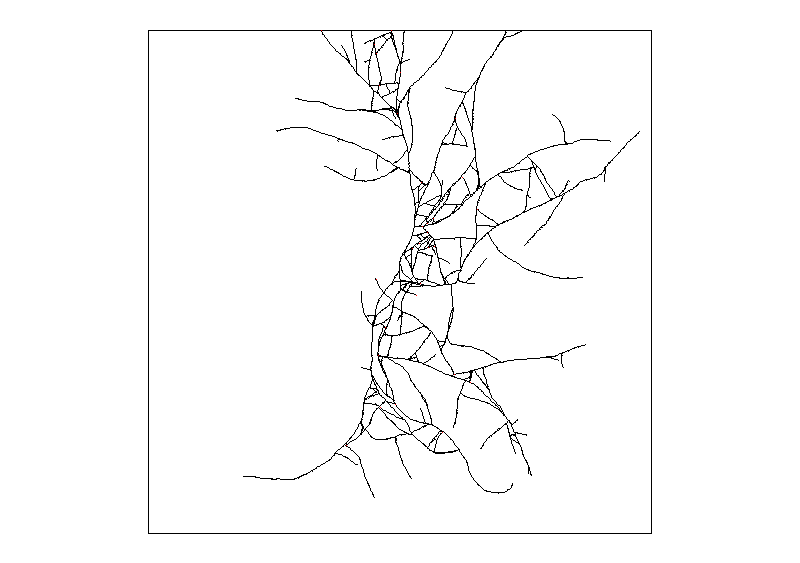}	\end{subfigure}
		\\ \vspace{0.3cm} \def\scaleFig{.20} \def\trimleft{0} \def\trimbottom{0} \def\trimright{0}  \def\trimtop{0}  \def\hDist{.4cm}
		\begin{subfigure}{\scaleFig \textwidth}
			\includegraphics[trim= \trimleft cm \trimbottom cm \trimright cm \trimtop cm,clip,width=\textwidth]{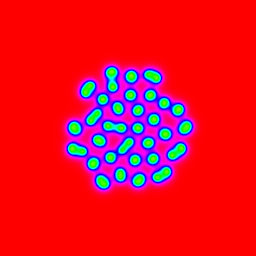}
			\caption{solitons (no feedback)}	
		\end{subfigure}
		\hspace{\hDist}	
		\begin{subfigure}{\scaleFig \textwidth}
			\includegraphics[trim= \trimleft cm \trimbottom cm \trimright cm \trimtop cm,clip,width=\textwidth]{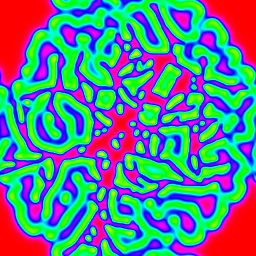}
			\caption{mazes}	
		\end{subfigure}
		\hspace{\hDist}	
		\begin{subfigure}{\scaleFig \textwidth}
			\includegraphics[trim= \trimleft cm \trimbottom cm \trimright cm \trimtop cm,clip,width= \textwidth]{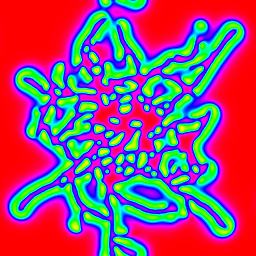}
			\caption{moving spots}
		\end{subfigure}
		\hspace{\hDist}	
		\begin{subfigure}{\scaleFig \textwidth}
			\includegraphics[trim= \trimleft cm \trimbottom cm \trimright cm \trimtop cm,clip,width= \textwidth]{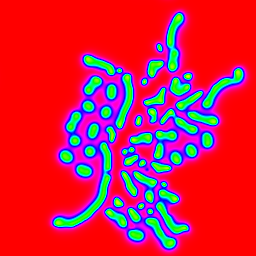}
			\caption{solitons}	\label{fig:exp_fb_neg_fk_solitons}
		\end{subfigure}
		\caption{Four experiments (columns). The first row indicates four final spatial networks and the second row the corresponding \gls{rd} layer.  }
		\label{fig:patterns}
	\end{figure}
	
	\begin{figure} [t] 	\def\scaleFig{0.49} 
		\centering
		\begin{subfigure}{\scaleFig \textwidth}
			\includegraphics[width=\textwidth]{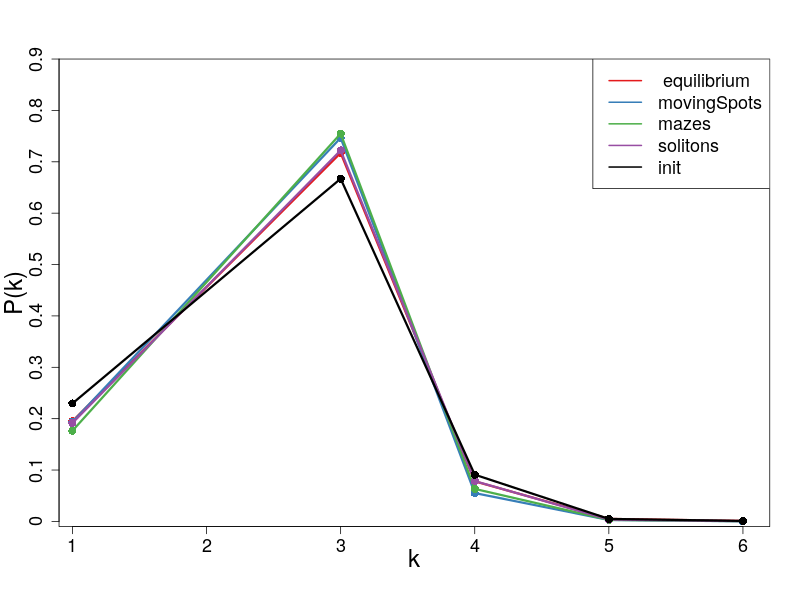}
		\end{subfigure}
		\begin{subfigure}{\scaleFig \textwidth}
			\includegraphics[width=\textwidth]{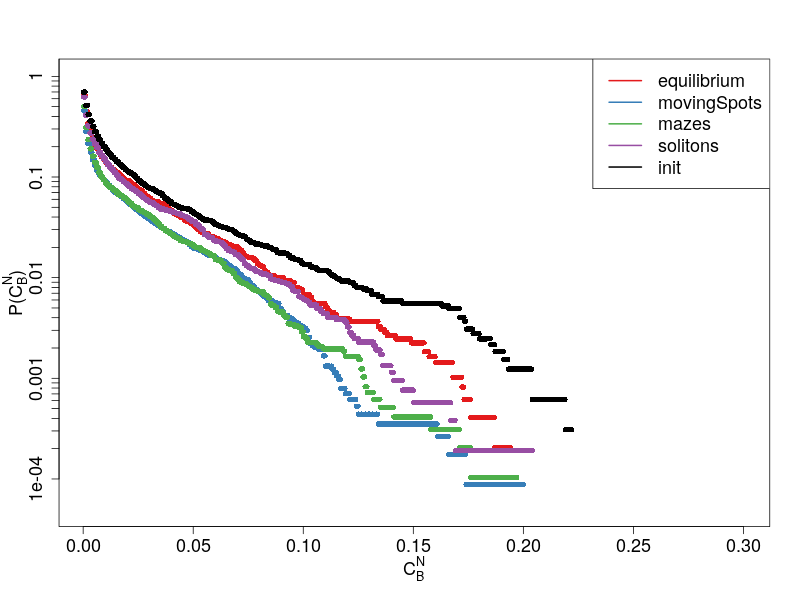}
		\end{subfigure}
		\caption{Vertex degree (left) and \acrlong{bc} (right) statistical distributions of Fécamp.
			Each panel compare the initial configuration and the configuration at the end of four simulation. }
		\label{fig:fecamp_deg_bc}
	\end{figure}
	
	Among different element that composed an urban system, we can suppose that some of them have a dominant role into the morphogenesis of the street network. 
	This latter is constrained by the spatial arrangement of such a form-producer elements and by exogenous factors (like the geomorphology or political decisions).  
	\myciteauthor{turing_chemical_1952} in his influential paper on chemical morphogenesis call them morphogens.
	The emergence of the network feeds back to the morphogens, affecting  their capacity to organize themselves. 
	
	We formalize those dynamics with a model composed by three layers surrounded by an environment (\cref{fig:concept}). 
	The first layer is cellular automata \parencite{wolfram_new_2002} constituted which simulates the spontaneous emergence of morphogens.  
	The concentration of morphogens updates in accordance with the \myciteauthor{gray_autocatalytic_1983} model, a particular model of the \gls{rd} theory. 
	The dynamic is expressed as a \gls{rd} system \parencite{turing_chemical_1952}, where two kinds of morphogens ($A$ and $B$) react and diffuse. 
	At a microscopic scale, $A$ catalyses its own production and also the production of $B$. 
	At the same time, $B$ inhibits the production of $A$.
	$B$ diffuses faster than $A$. 
	At a macroscopic scale, the \gls{rd} system can yield evolving patterns of concentration \parencite{pearson_complex_1993}.  
	The intermediate layer is a dynamic vector field that reflects the constraints generated by the \gls{rd} and by the environment. 
	Its role is to model the impact of those emerging patterns to the spatial network (the latest layer).  
	The network formation feeds back to its morphogenetic elements. 
	The mechanism consists to reduce the possibility of morphogens to cluster modifying parameters of the \gls{rd} layer.  
	The model is completed with a spatial environment, which aim to represent exogenous urban constraints for the street network (as like the geomorphology of the study case, socio-economical forces or policy decisions).	
	More details of the model and a complete overview about experiments can be founded in \autocite{tirico_morphogenesis_2020,tirico_morphogenesis_2019}. 
	
	\section{Results}	
	In \cref{fig:patterns} we present 4 simulations obtained with three pattern formation parameters. 
	The first simulation does not consider the feedback process.   
	Morphogens seem to bypass the graph. 
	Like a physical limit, a barrier defines a changing of behaviour of morphogens: on one side no organizations can form, and on the other side, morphogens create patterns according to the unperturbed pattern formation process.  
	This behaviour is unexpected, it diverges to classical simulation, making a wealthy of new and mixed patterns. 
	
	The second experiment concerns the application of the model in a real context: we study the street network morphogenesis of Fécamp town.  
	The environment integrate several morphogenetic aspects that had been considered during the simulation: the geomorphology (a second vector field was computed from an orographic map and combined with the main one), the build-up and the green area. 
	In green and built-up areas we reduced the probability of the network to develop. 
	In this way we integrate in the process, in a stylized manner, those political decisions that can impact the network evolution in a defined area.
	
	We measured the degree distribution (\cref{fig:fecamp_deg_bc}, left) and the \acrlong{bc} distribution (\cref{fig:fecamp_deg_bc}, right) \autocite{porta_network_2006} of the street network of nowadays and the networks after four simulations. 
	Simulated networks are obtained with four different patterns. 
	They conserve the main characteristics of the starting network with variations.
	Due to our decentralized approach, we observe that the growth of Fécamp town move toward an organic form, rich of tree-like networks and bifurcations.
	This process is captured by both distributions, where we observe an increment of vertices with degree 3 and a more hierchizated distribution of the \acrlong{bc}.   	
	
	\section{Discussion and conclusion} \label{sec:discussion}
	This work proposes a street network generator model. 
	Inspired by complexity theories, dynamics are completely decentralized and driven by feedback mechanisms between layers. 
	An external control can be eventually integrated by imposing over the environment some probabilities of growth. 
	The application of the model to study the evolution of the Fécamp town suggests that the growth of a street network is the result of an unpredictable combination of rate of growth, exogenous factors and feedback.
	The main properties of generated graphs are similar to real street networks. 
	In this context, morphogens can represent real activators and inhibitors of urban growth, e.g. population, economical factors and political actors. 
	We think the model may be helpful both to investigate urban growth and to support decisions of urban planners. 
	
	The fundamental brick of our stylized and general approach is that the morphogenesis of the street network is driven by morphogens. 
	They are spatially embedded, interact each others, are involved in competition/cooperation processes, move and may arranged with regularity. 
	We can find a correlation with those dynamics and population or economical actors of cities.
	We can suppose that it exists a relation between the concentration of people in a region of space and the existence of buildings and streets. 
	In this field, population could be considered for the street network as a morphogen because	it affects the network growth. 
	The spatial relation between the density of population and the street network is not in all situations ensured; it is still unpredictable, rarely synchronic, and often affected by a large amount of socio-economical and natural factors.
	Those preliminary observations open to many perspectives. 
	We start a process of validation. 
	More precisely, we plan to test our model in other urban contexts and validate our approach with a comparison between different real case studies. 
	
	\printbibliography
\end{document}